\documentclass[preprint,showpacs]{revtex4}

\usepackage{graphicx}
\usepackage{dcolumn}
\usepackage{amsmath}
\begin{document}

\title{A photonic bandgap resonator to facilitate GHz frequency conductivity experiments in pulsed magnetic fields. }

\author{R.~D.~McDonald}
\email[]{rmcd@lanl.gov}
\author{J.~Singleton.}
\author{P.~A.~Goddard}
\author{N.~Harrison}
\author{C.~H.~Mielke}
\affiliation{National High Magnetic Field Laboratory, LANL, MS-E536, Los Alamos, New Mexico 87545, USA.}

\date{\today}
\pacs{07.57.pt, 78.70.Gq,  84.40.-x,  07.55.Db}

\begin{abstract}
We describe instrumentation designed to perform millimeter-wave conductivity measurements in pulsed high magnetic fields at low temperatures. The main component of this system is an entirely non-metallic microwave resonator. The resonator utilizes periodic dielectric arrays (photonic bandgap structures) to confine the radiation, such that the resonant modes have a high Q-factor, and the system possesses sufficient sensitivity to measure small samples within the duration of a magnet pulse. As well as measuring the sample conductivity to probe Ôorbital physicsÕ in metallic systems, this technique can detect the sample permittivity and permeability allowing measurement of Ôspin physicsÕ in insulating systems. We demonstrate the system performance in pulsed magnetic fields with both electron paramagnetic resonance experiments and conductivity measurements of correlated electron systems.

\end{abstract}
\maketitle

\section{Introduction and overview}

Many effects of great scientific interest occur at high magnetic fields. These range from magnetic quantum oscillations, the measurement of which is an invaluable probe of the electronic band structure of metals~\cite{Shoenburg}, to magnetic-field-induced phase transitions such as the suppression of superconducting~\cite{Fedor} or charge-density-wave states \cite{RMcDPauli} and magnetic-field-induced quantum criticality~\cite{HarrisonURu2Si2}.
Such investigations of new materials and emergent phenomena continue to require ever higher magnetic fields. Quasi-static or DC magnets are currently limited to fields of 45~T, the generation of which requires 10s of MW of electrical power plus an equivalent cooling system; this of course represents a substantial investment in infrastructure~\cite{HerlachRepProgPhys}. By contrast, higher magnetic fields can be obtained for a significantly lower investment by using pulsed-field magnet systems. These have a much lower average power consumption and do not require extensive cooling system infrastructure~\cite{HerlachRepProgPhys}.  

Unfortunately, several compromises have to be made in order to take advantage of pulsed magnetic fields. The magnet bore is typically small compared to those of DC magnets, rapid measurement and data collection are required (the field pulse is typically 10s to 100s of milliseconds in duration) and the apparatus must be designed to cope with the large rates of change of field (typically $10^2-10^4$~Ts$^{-1}$) that can lead to inductive heating and very large impulsive forces in metallic components. Despite these technical challenges, many experimental techniques commonly used in condensed matter physics have been implemented in pulsed magnetic fields, including measurements of resistance, magnetization, sound speed, luminescence and heat capacity~\cite{OppsHighB}.  

Probing the excitations and ground state properties of correlated electron systems in high magnetic fields provides invaluable insight into their behavior. In many highly correlated-electron systems the energy scale of the interactions and excitations in high magnetic fields corresponds to light with a frequency of tens of GHz: 60~GHz $\approx$ 3~K $\approx$ $^1\slash_4$~meV. It is therefore desirable to extend GHz-frequency spectroscopy and conductivity measurements to fields that at present are only accessible by the use of pulsed magnets. In spite of this interest, very few far-infrared and microwave experiments have been carried out in pulsed fields, chiefly because these techniques traditionally rely upon metallic wave guides and resonant cavities that would be subject to enormous inductive heating and forces in the rapidly-changing magnetic fields. Such components are nevertheless crucial to the success of conventional DC-magnetic-field GHz-frequency apparatus, such as that used for electron paramagnetic resonance (EPR) or high-frequency conductivity experiments~\cite{MarijeJPCM}. The resonant cavity greatly improves the measurement sensitivity compared to a simple reflection or transmission geometry~\cite{MarijeJPCM}. Resonant cavities also facilitate the measurement of samples with dimensions that are small compared to the radiation wavelength. Waveguides enable measurements in environments ({\it e.g.} cryostats that fit into a magnet bore) that are of comparable size to the wavelength without the diffraction problems encountered with free space propagation. This aspect is particularly important at the frequencies to be considered in the present paper; 60~GHz radiation has a wavelength of approximately 5~mm in free space, comparable to the bore of many pulsed-magnet cryostats.

\begin{figure}
\centering
\includegraphics[height=5.4cm]{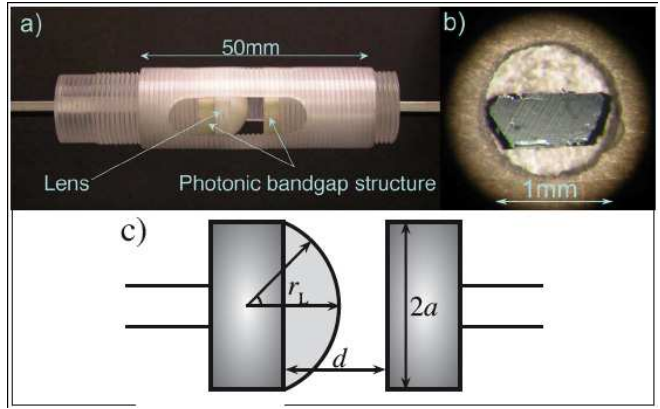}
\caption{a. The Fabry-Perot geometry photonic-bandgap resonator with waveguide coupling. b. A $\kappa$-(BEDT-TTF)$_{2}$Cu(SCN)$_{2}$ sample loaded in the dielectric resonator. This sample is approximately 1~mm long, 400~$\mu$m wide and 100~$\mu$m thick. c. a schematic of the resonator indicating the following dimensions: diameter of the optics ($2a$), radius of curvature of the lens ($r_{L}$) and the separation of the dielectric mirrors ($d$). (color online). } 
\label{DevicePic1}
\end{figure}

In this paper, we report solutions to many of the above issues that have prevented millimetre-wave experiments in pulsed fields; entirely non-metallic microwave resonant cavities that both avoid inductive heating and are of sufficiently small size to fit within a pulsed magnet. A typical resonator is shown in Figure~\ref{DevicePic1}; it relies upon photonic-band-gap structures to longitudinally confine the radiation and a central dielectric puck/lens structure to provide lateral confinement. As we shall describe below, the resonator has been used successfully in a number of pulsed-field experiments.

It is helpful to give an overview of the remainder of this paper as follows. Section~\ref{SSPBG} describes the photonic-bandgap structure, which consists of a periodic array of dielectrics, essentially a stack of quarter wave plates with sufficient refractive index contrast to possess a high reflectivity for a large range of frequencies and almost all angles of incidence. 

Section~\ref{SSRes} describes the two resonant cavity geometries; (i)~a Fabry-Perot resonator based on  the photonic- bandgap multilayers plus a plano-convex lens and (ii)~a cylindrical dielectric  cavity. For both types of cavity, coupling in the microwave power is via mono-moded rectangular waveguide. In the case of (i), the Fabry-Perot geometry, this generates resonant modes that are transversely-polarized with a Gaussian intensity profile. In the case of (ii), the cylindrical cavity formed by a dielectric puck sandwiched between the photonic bandgap multilayers, the resonances correspond to the transverse electric, (TE), transverse magnetic (TM) and hybrid electromagnetic (HEM) modes of the the dielectric cylinder, their relative strengths depending upon the exact coupling geometry.
In both cavity types, the longitudinal standing wave pattern is such that there is an integer number of half wavelengths between the multilayers with an electric field node and magnetic field antinode at the multilayer surface. This is an ideal geometry for many microwave experiments where it is desirable to have the oscillatory GHz magnetic field predominately orientated perpendicular to the applied magnetic field; examples include EPR and bulk conductivity measurements of layered materials~\cite{MarijeJPCM}. 

Discussions of data acquisition and interpretation are presented in Section~\ref{SSDataAq}. Loading the resonator with a sample perturbs the resonant frequency and Q-factor. Therefore, measuring the change in transmitted amplitude at resonance provides a very sensitive probe of the sample properties. Fast data acquisition at the frequencies in question is provided by a Millimeter-wave Vector Network Analyzer (MVNA) manufactured by AB Millimetre \cite{ABmm}; the MVNA was also used as a diagnostic instrument in the characterization of the various cavity designs and the multilayers~\cite{MarijeJPCM} (see Figures within Sections~\ref{SSPBG} and \ref{SSRes}.  Ideally the sample provides a small perturbation to the resonator. The size of this perturbation, and hence the sensitivity to the change in sample properties, can be controlled via the sampleÕs radial location in the resonator. For low-loss or small samples, the perturbation is optimal for the sample located centrally. This is achieved by mounting the sample in a thin disk of the lens material clamped between the lens and one of the dielectric mirrors (see Figure~\ref{DevicePic1}b). 

Section~\ref{SSTests} illustrates the system performance in pulsed magnetic field. The apparatus is currently designed to run in a 24~mm bore, 50~Tesla pulsed magnet equipped with a flow cryostat providing stable temperatures between room temperature and 2~Kelvin. The rise time of the magnetic field is 8~ms with a current of 17~kA at peak field, obtained by discharging a 16~mF capacitor bank from a voltage of 7.6~kV (0.46~MJ of energy). Illustrative examples include measurements of the magneto-conductivity of the organic superconductor $\kappa$-(BEDT-TTF)$_2$Cu(SCN)$_2$ at a frequency of 64~GHz and a temperature of 2.1~Kelvin. Under these conditions the quantum oscillations corresponding to the $\alpha$ and $\beta$ orbits about the Fermi surface~\cite{SingletonRepProgPhys} are clearly visible. A summary and conclusions are given in Section~\ref{summary}.

\section{The photonic bandgap structures.} \label{SSPBG}
\begin{figure}
\centering
\includegraphics[height=7cm]{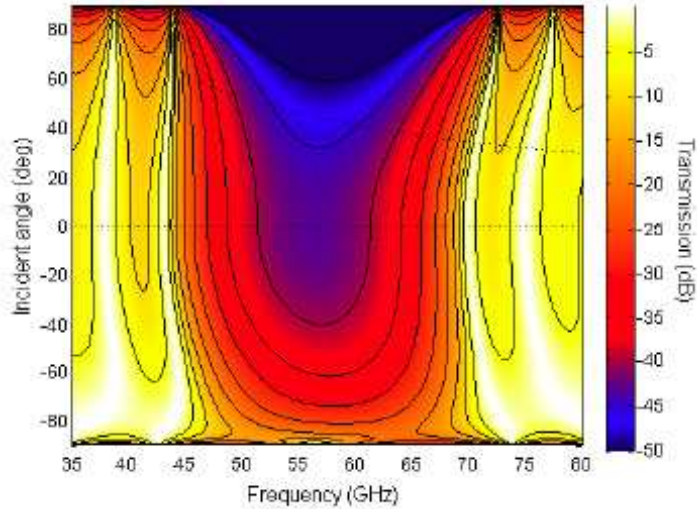}
\caption{The calculated transmission through the 1D-photonic bandgap structure described in the text as a function of frequency and angle of incidence. Positive angles correspond to TE or s-polarization and negative angles to TM or p-polarization. The lower dashed line is normal incidence and the upper the effective angle of incidence for the TE$_{10}$ mode of V-band rectangular waveguide (color online).} 
\label{PBGsurf}
\end{figure}

The use of two multilayer dielectric structures to provide longitudinal confinement of the radiation is common to all of the variants of the non-metallic resonant cavity described below. These structures are essentially quarter-wave-plate stacks or distributed Bragg reflectors. However, it is convenient to both treat and refer to them as {\it photonic-bandgap structures} for several reasons. Firstly, in order to be efficient mirrors for this application, they must not only be highly reflective at normal incidence, but also for a wide range of angles of incidence. This can be realized by considering what are effectively modes of a lattice, {\it i.e.} decomposing the standing wave solution of the resonant cavity into counter-propagating traveling waves and examining the component of wavevector or momentum parallel to the dielectric surface~\cite{Pozar}. For example, the upper dashed line in Figure~\ref{PBGsurf} corresponds to the effective angle of incidence for light propagating in the TE$_{10}$ mode of rectangular V-band waveguide. Note that as the low-frequency cutoff is approached the effective angle of incidence diverges from normal. 
Secondly, although the thickness of the lens in the Fabry-Perot geometry enables a quasi-geometrical optics approach to be applied to the analysis of its modes, the sub half wavelength optical thickness of the air gap surrounding the puck in the cylindrical geometry dictates that when considered as a whole,  complete resonant cavity (end reflectors plus puck) should be regarded as a photonic-bandgap structure with a central resonant defect. 

A high reflectivity for a wide range of incidence angles is achieved by using materials with a high dielectric contrast. This has the added advantage in that the frequency width of the highly reflective region is a substantial fraction of the center frequency. Figure~\ref{PBGsurf} shows the calculated transmission spectrum for a five and a half period dielectric stack. Each period consists of $380~\mu$m of Alumina \cite{alumina}, with a relative permittivity of 9.9, and $250~\mu$m of Zirconium Tin titanate \cite{ZTT}, with a relative permittivity of 36.5. The structure is terminated on both sides with the higher dielectric constant material to optimize the reflectivity. The photonic bandgap can be tuned to the frequency range of interest simply by selecting materials with the desired dielectric constant and thickness. The calculation of the reflection, transmission and absorption coefficients of one-dimensional multilayer systems were performed using a transfer-matrix formulation of the Frensel equations \cite{P&P}.

\begin{figure}
\centering
\includegraphics[height=6.2cm]{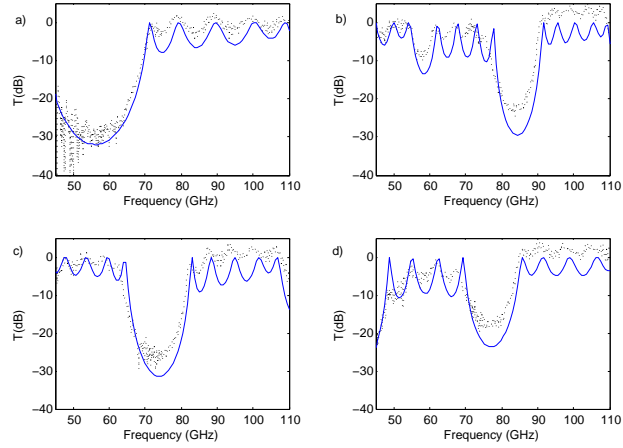}
\caption{The measured (dashed line) and calculated (solid line) transmission through four different $4^{1}/_{2}$ multilayer structures illustrating the tuneability of the bandgap with layer thickness. Layer thicknesses and materials are as follows: a) 240~$\mu$m Zirconium Tin titanate 380~$\mu$m Alumina, b)  480~$\mu$m Zirconium Tin titanate 760~$\mu$m Alumina, c)  480~$\mu$m Zirconium Tin titanate 380~$\mu$m Alumina, d)  240~$\mu$m Zirconium Tin titanate 760~$\mu$m Alumina (note the logarithmic scale: -20~dB $ = 1\%$) (color online).} 
\label{PBGTune}
\end{figure}

To optimize the Q-factor of the system, the dielectric materials are chosen for their low loss tangent: in this example, $2\times10^{-4}$ for Alumina and $2.5\times10^{-5}$  for Zirconium Tin Titanate. 
Figure~\ref{PBGTune} shows the calculated and measured transmission spectrum of several structures fabricated from Alumina and Zirconium Tin titanate, where the thickness of each has been used to tune the bandgap to the frequency range of interest. The free-space transmission was measured at room temperature using the MVNA  as source and detector~\cite{ABmm,MarijeJPCM} (see also Section~\ref{SSDataAq}) with the radiation emitted from and collected by rectangular waveguide. The high frequency oscillations that appear as noise on the data but not the simulation are due to standing waves in the system; their amplitude was minimized by the use of ferrite directional couplers. Again the simulations are calculated via a transfer-matrix formulation of the Frensel equations~\cite{P&P}. 

For mechanical stability the dielectric stacks are laminated using Stycast~1266 two-part epoxy. Stycast~1266 was chosen for its mechanical resistance to thermal cycling and low viscosity prior to curing. The low viscosity is vital to ensure the resulting optical thickness of the epoxy is orders of magnitude smaller than that of the dielectrics, so as not to impact the performance of the structure. In practice, the cured epoxy is less than 1$\mu$m thick, and has a negligible effect.

The optimum number of periods of the photonic bandgap structure depends upon the resonator geometry: consequently, discussion of this topic is covered in the following section. 

\section{Resonant cavity geometry and coupling.} \label{SSRes}
\subsection{General considerations}
\begin{figure}
\centering
\includegraphics[height=11cm]{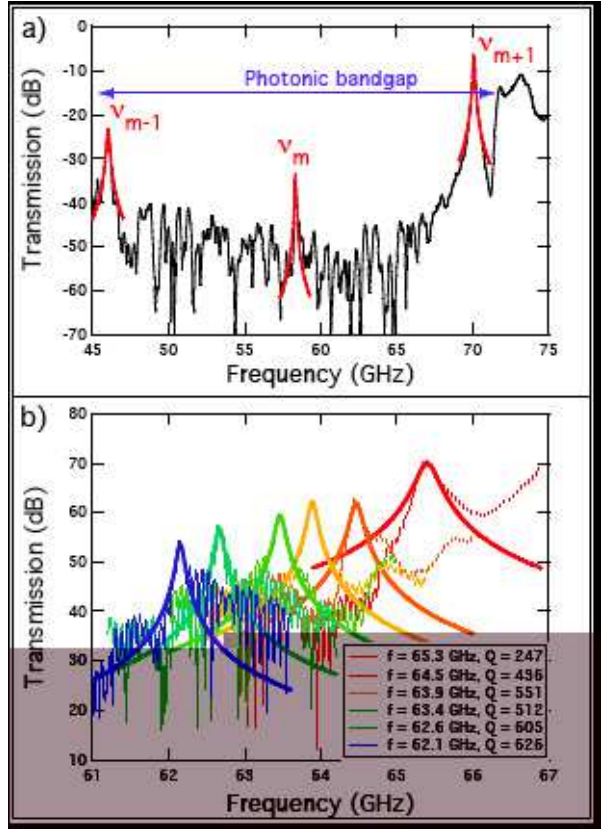}
\caption{a. The room temperature transmission spectrum of the Fabry-Perot style resonator, illustrating the low-transmission photonic bandgap region and well separated resonant modes labeled by their longitudinal mode index ($m = 4$). 
b. Tuning the resonance in the region of the band edge, note that the increased coupling is accompanied by an increase in the width of the resonance, a degradation of the Q-factor (color online) .} 
\label{ResFig1}
\end{figure}

Two distinct geometries of resonant cavity employing the dielectric multilayers have been developed and used for pulsed field measurements: a Fabry-Perot style resonator and a cylindrical dielectric cavity resonator. Rather than using reflection coupling, it is preferable to measure both geometries of resonant cavity in transmission such that a resonance appears as a sharp peak in intensity, orders of magnitude larger than the background signal (see Figure~\ref{ResFig1})~\cite{Pozar}. The coupling of the microwave power to the resonant cavity by terminating the waveguide with the resonatorÕs multilayer structure has an advantage over aperture coupling in that it is relatively frequency independent: within the frequency range of the photonic bandgap only a decaying wave penetrates the resonator and critical coupling can be obtained by varying the number of dielectric layers (see Figure~\ref{QfacOsc1}a) or by tuning the resonant frequency to the edge of the photonic bandgap (see Figure~\ref{ResFig1}b). In practice it is usually preferable to slightly overcouple the resonator, at the expense of the Q-factor, so as to maximize the difference between transmission through the resonator and the background microwave leak (the ``leak signal") that bypasses the resonator. Note from Figure~\ref{ResFig1}a that at room temperature the transmission is of the order of -50 dB down at mid band gap and that the resonant modes are of the order of 20 dB above this.  For both resonant structures the insertion loss is dominated by the number of  periods of the dielectric multilayers, for example, as can be see from Figure~\ref{ResFig1}, away from resonance the insertion loss is between -40 and -60~dB throughout the photonic bandgap region. At low temperature both the Q-factor and ratio of resonant amplitude to background transmission improve due to the reduction of dielectric losses. 

The internal diameter of the Helium flow cryostat for use in the 24~mm bore, 50~T pulsed magnet system is 17.5~mm. Allowing space for the flow of Helium and the body of the resonator, which provides mechanical support, the maximum diameter of the optics is 14~mm (the diameter of the photonic-bandgap structures and lens in Figure~\ref{DevicePic1}). When performing measurements at V-band frequencies (45-75~GHz) with 14~mm diameter optics, ${5^{1}\slash_{2}}$ periods of alumina and zirconium tin titanate provides the optimal compromise between coupling and Q-factor. For larger diameter optics the leak signal is reduced, such that it is possible to use a larger number of periods to increase the reflectivity and hence Q-factor, without compromising the dynamic range of the resonance.

Thin-walled stainless steel waveguide has a low enough conductivity and cross-section perpendicular to the applied field to be used in a 50~T pulsed magnet with a rise time of 8~ms, without significant heating. For the resonator to be used in a pulsed magnet with a significantly larger rate of change of field, the stainless steel waveguide would have to be replaced by dielectric waveguide near the field center. Measurements indicate that this results in a slightly larger microwave leak signal but does not change the overall performance of the resonator~\cite{PBGtobepublished}.

\subsection{The Fabry-Perot geometry.}
Fabry-Perot or open resonators usually rely upon the use of curved mirrors to confine the radiation laterally within the cavity, thus maintaining a satisfactory Q-factor~\cite{Kogelnik&Li,Clarke}. Unfortunately, producing a spherically curved multilayer dielectric stack presents significant technological challenges. As a result, the Fabry-Perot geometry resonator uses a plano-convex spherical lens to confine the radiation laterally. However, in addition to the desired refraction (leading to confinement), reflection from the lens surface will cause unwanted scattering of light out of the resonator. A planar-convex lens has an advantage in that there is only one refractive surface when its planar side is in direct contact with one of the dielectric mirrors. This is an important consideration, because, as we show below, even for low refractive index lenses the unwanted reflection will be $\sim 2$~\%, by far the dominant contribution to degradation of the Q-factor. 

\begin{figure}
\centering
\includegraphics[height=11cm]{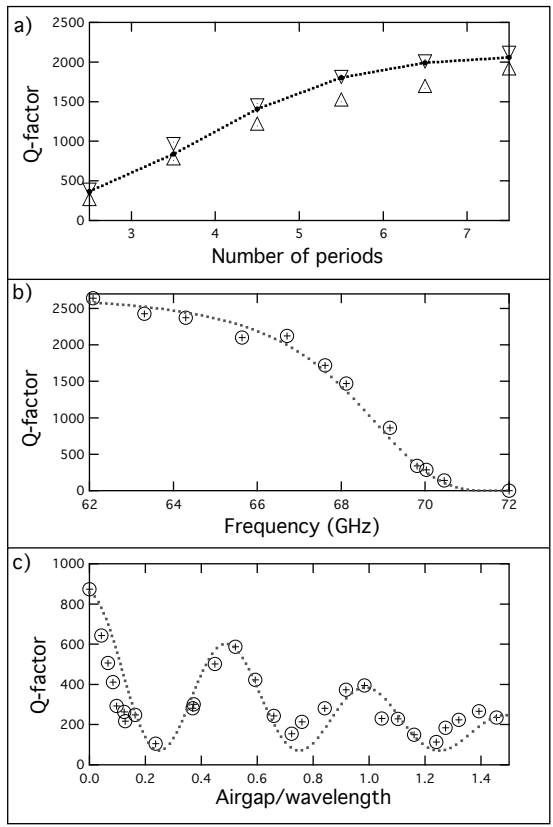}
\caption{a. The Q-factor of a 64~GHz mode as a function of of the number of periods of the dielectric multilayer. The triangles mark the range of observed Q-factors and the solid points and dashed line are the simulation described in the text. 
b. The Q-factor of a single mode as it is tuned to the edge of the photonic band gap, the crosses are data and the line the simulation described in the text. The data in both a. and b. were measured at room temperature using Fabry-Perot style resonator with one dielectric multilayer and one spherically curved copper mirror. The aperture coupling through the copper mirror limits its effective reflectivity to $\approx$~99.75$\%$ and is responsible for the saturation of the Q-factor in both a. and b. 
c. Oscillations in the Q-factor of the Fabry-Perot style photonic bandgap resonator. The data (crosses) correspond to multiple modes of the resonator, tuned to near the center of the photonic bandgap via the length of the air-gap between the lens and one of the dielectric mirrors. The simulation (dashed line) uses a lens reflectivity, $\mathcal{R}_L$, of 95$\%$ and a dielectric mirror reflectivity, $\mathcal{R}_M$, of 99.7$\%$ to account for the observed amplitude of oscillation, the period and decay are predicted via the lens geometry as described in the text.}
\label{QfacOsc1}
\end{figure}

Figures~\ref{QfacOsc1}a and b illustrate how the Q-factor of a resonator utilizing the dielectric multilayer structures is dependent upon their reflectivity. In Figure~\ref{QfacOsc1}a the multilayer reflectivity is tuned via the number of periods and in Figure~\ref{QfacOsc1}b by tuning the resonance towards the edge of the photonic bandgap. The ability to model this behavior provides an important starting point to understanding the performance of the dielectric resonator as a whole. To first approximation the Q-factor can be estimated using~\cite{Kogelnik&Li,Clarke}, 
\begin{equation}
Q = \frac{m\pi \mathcal{R}^{\frac{1}{2}}}{1-\mathcal{R}},
\label{Qequ1}
\end{equation}
where $m$ is the index of the longitudinal mode and $\mathcal{R}$ is the effective reflectivity of the multilayer structure, $\mathcal{R}_{\rm M}$, scaled by one minus the reflectivity of the lens, $1-\mathcal{R}_{\rm L}$, and the beam confinement fraction $\Phi_{\rm Frac}$, 
\begin{equation}
\mathcal{R} = \mathcal{R}_{\rm M}(1-\mathcal{R}_{\rm L})\Phi_{\rm Frac}.
\label{Requ1}
\end{equation}
$\Phi_{\rm Frac}$ is given by the surface integral of the intensity profile from the resonator center to the radius of the optics, $a$ \cite{P&P, Kogelnik&Li}.  For the fundamental gaussian mode
\begin{equation}
\Phi_{\rm Frac} = 1 - e^{-\frac{2a^{2}}{w^{2}}}.
\end{equation}
Here, $w$ is the beam radius, defined (in the usual manner) as the radius at which the amplitude has fallen to $^{1}\slash_{e}$ of its peak value~\cite{Kogelnik&Li,Self}. 
For an etalon with one planar mirror and one concave mirror with a radius of curvature $r_{\rm M}$, the maximum beam radius occurs at the surface of the concave mirror and is given by, 
\begin{equation}
w^{2} = \frac{\lambda r_{\rm M}}{\pi} \bigg\slash{ \sqrt{\frac{r_{\rm M}}{d}-1}},
\label{BeamRadiusEqu}
\end{equation}
where $\lambda$ is the wavelength of radiation, and $d$ the mirror separation \cite{Kogelnik&Li}. 
For the case of planar mirrors and a planar-convex lens, the maximum beam radius occurs at the refracting surface of the lens. If the mirror radius of curvature is replaced by the focal length of the lens, by symmetry \cite{Kogelnik&Li}, the same expression (Equation~\ref{BeamRadiusEqu}) can be used to estimate the maximum beam radius, and hence the confinement factor. The focal length of a single surface lens, of refracticve index $n_{\rm L}$, and with a radius of curvature $r_{\rm L}$ is given by 
\begin{equation}
f = r_{\rm L}\left( \frac{n_{\rm L}}{n_{\rm L}-1}\right).
\end{equation} 
Thus by using suitable dimensions and material properties for a 65~GHz resonator ($a$ = 7~mm, $d$ = 6~mm, $r_{\rm L}$ = 7.5~mm and $n_{\rm L}$ = 1.6) it can be seen, via Equation~\ref{Qequ1}, that the beam confinement contribution limits the Q-factor to $\approx10^{3}$. By comparison, approximating scattering losses from the surface of the lens as normal incidence reflection, yields
\begin{equation}
1-\mathcal{R}_{\rm L} = 1- \frac{\left(1-n_{\rm L}\right)^{2}}{\left(1+n_{\rm L}\right)^{2}} \approx 95\%,
\end{equation}
which for the same geometry as above limits the Q-factor to $\approx 10^{2}$.
Approximating the reflection from the lens in this manner is exceedingly crude, however, and in reality only serves to highlight how important a consideration this is when optimizing the Q-factor. This simple intensity approximation ignores the fact that centrally the tangent of the curved surface of the lens is perpendicular to the axis of the resonator, such that light reflecting from it is not scattered out of the resonator, but gains a phase shift relative to the transmitted light that is reflected from the dielectric mirror. The amplitude approximation therefore represents the worst case scenario, in which the reflections are out of phase. By adjusting the relative thickness of the lens and the resonator length, such that both are an integer number of half wavelengths at resonance, the phases of the reflections are matched. This constructive interference greatly enhances the resulting Q-factor. Figure~\ref{QfacOsc1}c illustrates the periodic nature of the Q-factor as the length of the resonator is changed. The decay of the Q-factor with increasing mirror separation is due to the loss of beam confinement. The data in Figure~\ref{QfacOsc1}c are measured with the resonance tuned to near the middle of the photonic bandgap, so as to avoid the influence of the frequency dependence of the dielectric mirror reflectivity, $\mathcal{R}_M(\omega)$: at mid-bandgap, the dielectric mirror reflectivity is far from being the limiting contribution to the Q-factor (see Figure~\ref{QfacOsc1}b). The accuracy with which the simulations in Figures 5a, b and c reproduce the trend in observed Q-factor indicate that the quasi-geometrical optics approach we employ to model the performance of the Fabry-Perot geometry resonator is appropriate.

The most trivial way to ensure the constructive interference condition is met is to make the resonator length equal to the lens thickness. Clamping the lens between the multilayers also reduces the effect of vibrational transients during a magnet pulse. Another consideration in selecting the dielectric material for the lens is its thermal properties: because this apparatus is designed to operate over a wide range of temperatures (between room temperature and 2~K) differential thermal contraction between the resonator body, and dielectric components is highly undesirable. To minimize this issue, the same material is used for the resonator body and lens.  Rexolite 1422 is an ideal material, as it is a  thermally stable plastic, with a low dielectric loss at GHz frequencies~\cite{Pozar}. It is also readily machinable.

When measuring anisotropic materials it is highly desirable to be able to orient the sample precisely with respect to the applied magnetic field and the microwave-frequency electric and magnetic fields~\cite{MarijeJPCM,HillRSI}. This requires that the resonator has a distinct polarization, which is achieved  by using a non birefringent lens and coupling to the resonator with mono-moded rectangular waveguide. At room temperature the resonant amplitude with the polarization of the input and output waveguides parallel is more than two orders of magnitude greater than with the polarizations crossed. This indicates that the transverse field components within the resonator are polarized to higher than 99$\%$. 

\subsection{The cylindrical puck geometry.}
Microwave radiation is significantly less well confined within a solid dielectric cavity than within its hollow metallic counterpart. As a result, perturbation of the fringe fields is an important consideration in the optimization of the Q-factor, especially when designing a resonator to operate in a confined environment such as a pulsed magnet and cryostat. 

For a resonator with metallic walls, the decay of the microwave field outside the cavity is determined by the skin depth of the metal: for copper at 60~GHz and room temperature the skin depth is $\approx$ 0.25~$\mu$m. For a dielectric cavity in free space the decay of the microwave field outside the resonator is determined by the contrast between the dielectric constant of the resonator and its surroundings~\cite{Pozar,Pospieszalski}. For a moderate dielectric-constant contrast, this decay length is of the same order as the free space wavelength. This has the advantage that placing a sample in the proximity of the resonator provides tunable coupling to the microwave field. However, the resonance will also be strongly perturbed by its surroundings, {\it i.e.} microwave power will be lost to the cryostat and magnet, thus reducing the Q-factor and contributing an unwanted temperature- and magnetic-field dependence of the resonance. 

\begin{figure}
\centering
\includegraphics[height=4.5cm]{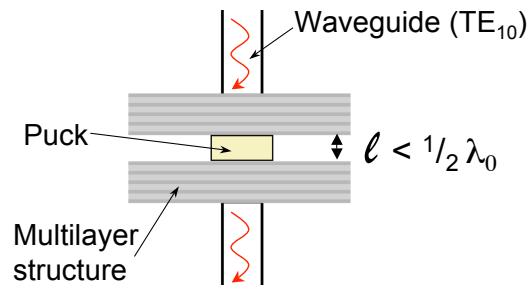}
\caption{Schematic of the cylindrical puck geometry resonator (color online).} 
\label{Puck1}
\end{figure}

Analogously to the Fabry-Perot geometry dielectric resonator described above, the puck geometry resonator utilizes the dielectric multilayer structures to increase the longitudinal confinement of the microwave field. The necessary radial confinement for operation in the bore of a pulsed magnet is obtained by reducing the length of the dielectric puck to less than half the free space wavelength (see Figure~\ref{Puck1}).  Without the central dielectric, the multilayer-air gap-multilayer structure as a whole possesses a photonic bandgap (consider for example, the ideal case in which the length of the air gap is a quarter of the free space wavelength). The dielectric puck can thus be regarded as a resonant defect in a photonic bandgap structure. As such, coupling to the evanescent tail of resonant mode is suppressed. 


The modes of this resonator resemble those of a dielectric puck resonator where the non-radiant condition is ensured by parallel conducting end plates \cite{Pospieszalski,Frezza,Kobayashi,Kajfez}. The frequency and field distribution of the resonant modes may thus be calculated in a similar fashion \cite{Pospieszalski,Frezza,Kobayashi,Kajfez} by making the approximation that the highly reflecting multilayer structures, capping both ends of the puck, provide perfect electric-wall boundary conditions. The imperfect magnetic-wall boundary condition at the circumference of the puck insures that, for modes that are not azimuthally symmetric, neither the oscillating electric or magnetic field components are purely transverse to the axis of the resonator. In other words only modes with no azimuthal field variation ($n = 0$) are transverse electric, TE$_{nml}$, or magnetic, TM$_{nml}$, where $m$, and $l$, correspond to the number of radial and longitudinal variations in the sign of the fields \cite{Kajfez}. Modes with azimuthal field variation ($n > 0$) are categorized as hybrid electromagnetic, HEM$_{nml}$ \cite{IEEE}. HEM modes with an even radial index are considered quasi-TE due to the greater component of transverse electric field and conversely those with odd radial index, quasi-TM.  The degree to which each mode is quasi-TE vs TM can be parameterized by the ratio of the microwave power in the TE and TM components of their fields ${\rm P^{TE}_{{\it nml}}/P^{TM}_{{\it nml}}}$ \cite{Kajfez}. For the two most dominant modes of this structure  (HEM$_{111}$ and HEM$_{121}$) this ratio is calculated to be ${\rm P^{TE}_{111}/P^{TM}_{111}} = 0.5$ and ${\rm P^{TE}_{121}/P^{TM}_{121}} = 5.0$. 

\begin{figure}
\centering
\includegraphics[height=14cm]{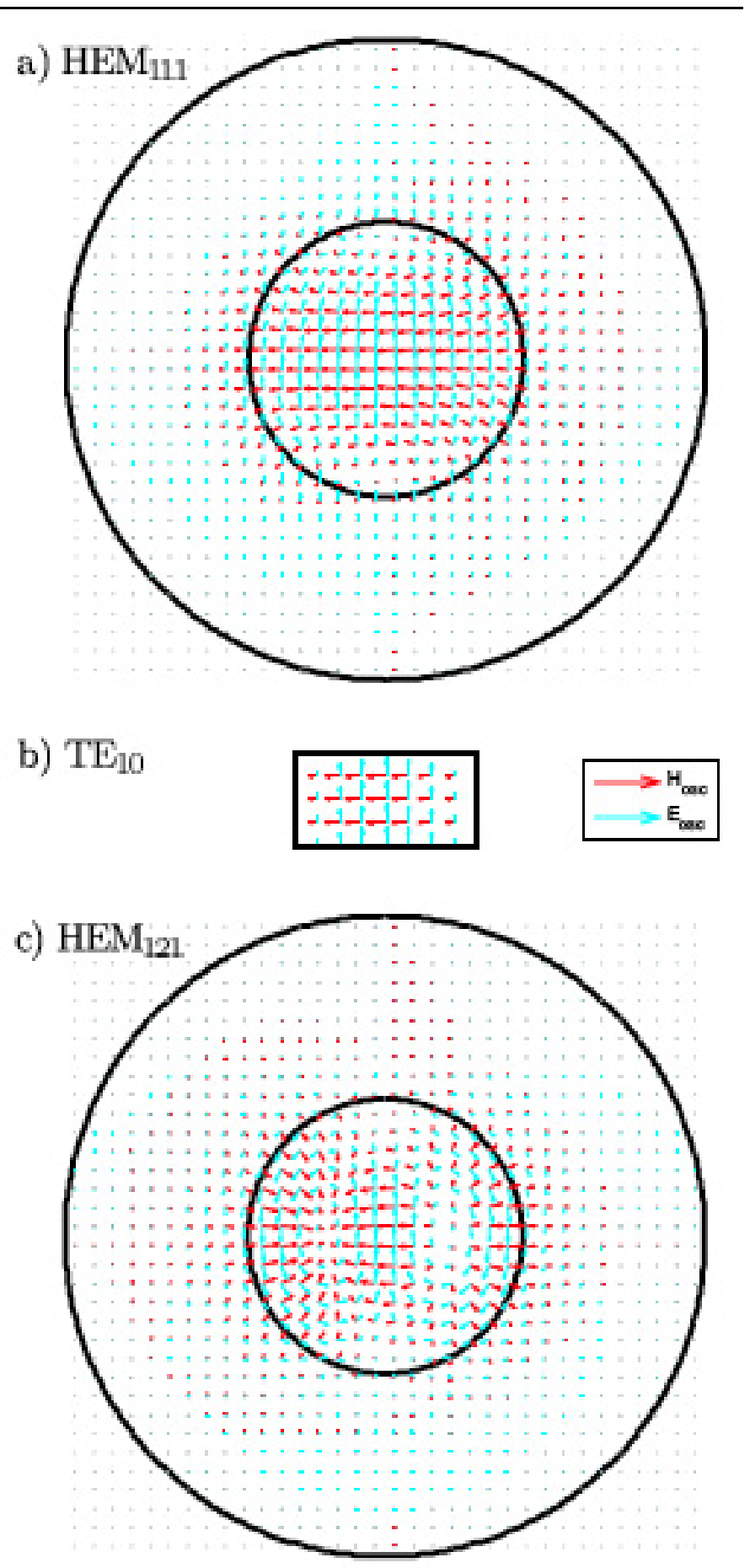}
\caption{The radial and azimuthal field patterns for the HEM$_{111}$ mode of the puck resonator (a), the TE$_{10}$ mode pattern for the rectangular waveguide (b) and the HEM$_{121}$ mode of the puck resonator (c).  The inner circle in a and c represents the circumference of the dielectric puck (in this case chosen to be 6~mm in diameter) and the outer circle, the circumference of the resonator (14~mm in diameter). The waveguide and resonator are drawn to scale so as to illustrate the similarities between the magnetic field distribution when aligned centrally. } 
\label{Modes1}
\end{figure}

\begin{figure}
\centering
\includegraphics[height=6cm]{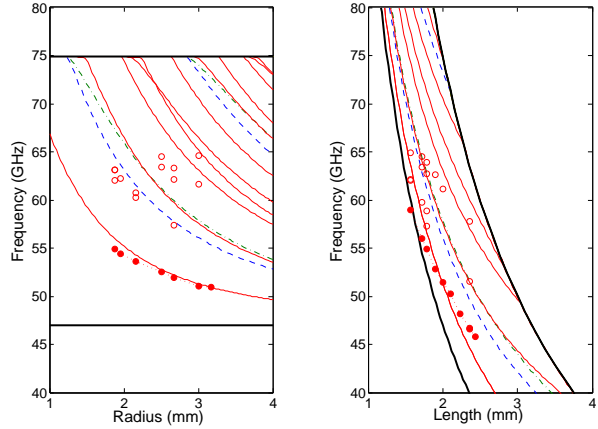}
\caption{The observed resonant frequency of the HEM$_{111} $ mode (the solid circles) as a function of the cavity radius for a 2~mm long cavity (a) and cavity length for a 6~mm diameter cavity (b). The frequency of other observed modes are hollow circles. The calculated frequency of the TE and TM and HEM modes are the dashed, dot-dashed and thin solid lines respectively.  The thick solid lines represents the cut off frequencies: above the upper cutoff the puck is not shorter than half the free space wavelength and below the lower cutoff the puck is shorter than the dielectric medium wavelength. Note that the strong agreement between the measured and calculated frequency of the HEM$_{111}$ is an indication of the degree to which the dielectric multilayers and the geometry suppress the fringe fields (color online).} 
\label{FcylFig1}
\end{figure}

\begin{figure}
\centering
\includegraphics[height=3.7cm]{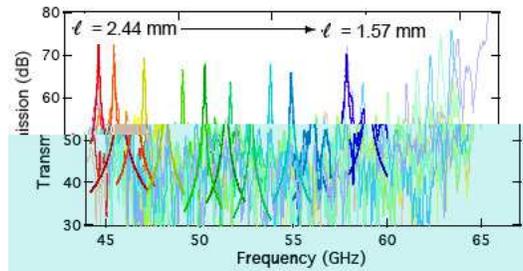}
\caption{Transmission spectra illustrating tuning of the HEM$_{111}$ mode of the dielectric-puck geometry photonic-bandgap resonator. The puck is a 6~mm diameter Rexolite~1422 cylinder of various lengths between 1.57~mm and 2.44~mm. The dielectric multilayers are 5$^1\slash_2$ periods of Zirconium Tin titanate and Alumina which are 14~mm in diameter (color online).}
\label{TwCylFig1}
\end{figure}

To understand why coupling to the HEM$_{111}$ and HEM$_{121}$ modes predominates we must examine the field patterns of these modes and the coupling waveguide. Applying the above boundary conditions yields the field distribution for the HEM$_{111}$ and HEM$_{121}$ modes of the resonator shown in Figures~\ref{Modes1}.  As can be seen from Figures~\ref{Modes1}a and c, the fields are concentrated within the dielectric puck (the inner circle) and decay towards the radius of optics (the outer circle). 
Coupling microwave power from mono-moded (TE$_{10}$) rectangular waveguide with the dielectric cylinder on the same axis, predominantly excites the HEM$_{11l}$ modes of the cylinder. This is because, for these modes, the field patterns on either side of the dielectric multilayer structure are well matched: both modes have a single-component, transverse-oscillating magnetic field antinode at the surface (see Figure~\ref{Modes1}). For a cylinder of suitable radius with a moderate dielectric constant, such as a 3~mm radius cylinder made of Rexolite 1422, the condition that the length of the puck is less than a half free-space wavelength only holds for the HEM$_{11l}$ mode with $l = 1$ (see Figure~\ref{FcylFig1}a). The frequency of the HEM$_{111}$ mode can be tuned throughout the frequency range of the photonic bandgap, by adjusting the length, radius and dielectric constant of the puck (see Figures~\ref{FcylFig1}a,b and \ref{TwCylFig1}). 
The following tabulates the calculated mode frequencies for a 3~mm radius, 2~mm long Rexolite puck so that the modes corresponding to the frequencies plotted in Figure~\ref{FcylFig1} may be identified.  

\begin{figure}
\centering
\includegraphics[height=5cm]{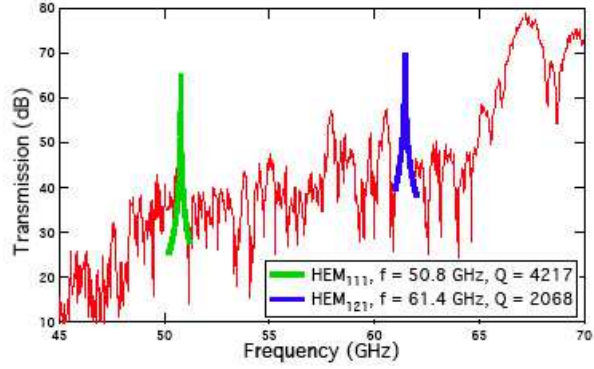}
\caption{The transmission spectrum of the unloaded puck resonator measured at 2.1~K. The puck is a Rexolite~1422 cylinder 2~mm long and 6~mm in diameter and the dielectric multilayers are 5$^1\slash_2$ periods of Zirconium Tin titanate and Alumina which are 14~mm in diameter. Both the HEM$_{111}$ and HEM$_{121}$ modes are clearly visible (color online). }
\label{PuckLowTres1}
\end{figure}

Other modes excited in this frequency range (for example HEM$_{121}$) appear weaker in the transmission spectrum due to a greater degree of mismatch between the field distribution in the waveguide and cylinder (see Figure~\ref{Modes1}b). For the unloaded resonator, the Q-factor of the HEM$_{111}$ mode exceeds 4000 at 2.1 K (see Figure~\ref{PuckLowTres1}). 

A potential advantage of the non-metallic resonators we have described in this section over conventional metallic resonators is the thermal stability of their performance. Unlike metallic resonators where the Q-factor is dominated by the surface resistivity and hence changes by more than an order of magnitude between room and cryogenic temperatures \cite{Pozar} the change in Q-factor of our resonator is less that a factor of two over the same temperature range. This weak temperature dependence is due to the reduction of the dielectric loss tangents at low temperature \cite{Lamb}.
 
\section{Data Acquisition.} \label{SSDataAq}
The GHz-frequency measurements reported here were performed using a Millimeter-wave Vector Network Analyzer (MVNA) manufactured by AB Millimetre \cite{ABmm}. The details of its use and operation are described elsewhere \cite{MarijeJPCM,HillRSI}. Here, we give a brief overview of the properties that make the system ideal for our purposes; {\it i.e.} the MVNA is a vector analyzer with a large dynamic range ($>$120~dB below 100~GHz) and its response time is fast enough to make measurements in pulsed magnetic fields with millisecond durations.

The MVNA consists of a source and detector of radiation in the frequency range from 8~GHz to beyond 1~THz (beyond 350~GHz Gunn diode extensions are required to boost the power in the higher harmonics). It is a super heterodyne detection system, {\it i.e.} it operates by mixing harmonics of two phase-locked oscillators of different frequency. In this manner, the phase and amplitude of millimeter-wavelength radiation can be measured using RF electronics. 

\begin{figure}
\centering
\includegraphics[height=5cm]{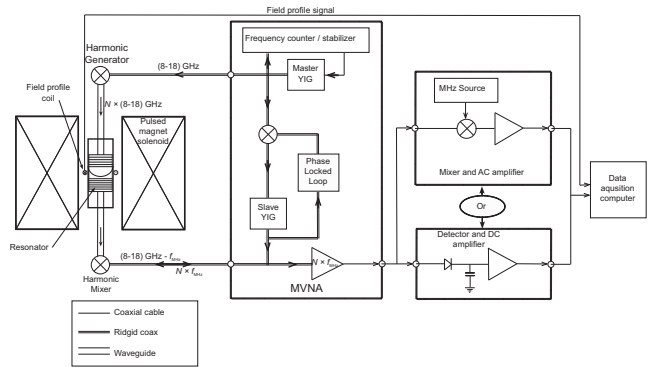}
\caption{Flow diagram of the pulsed field microwave detection system including the MVNA.}
\label{MVNAflow}
\end{figure}

The MVNA's  sources are YIG (Yttrium-Iron-Garnet) oscillators, which are continuously tunable between 8 and 18~GHz (the fundamental harmonic of the MVNA). One of the YIG oscillators (the master) is controlled by the operator, the second (the slave) is phase locked to the first and follows with a fixed difference frequency, $f_{\rm MHz}$.  The difference frequency is chosen depending upon the desired operating harmonic, $N$, such that $N \times f_{\rm MHz}$ corresponds to a frequency for which the vector receiver is optimized ($\approx$~9~MHz, $\approx$~34~MHz or $\approx$~59~MHz). Harmonics of the YIG oscillators are generated by Schottky diodes: the oscillating current from the master YIG drives a Schottky diode (the Harmonic Generator (HG)), which acts as a transducer emitting electromagnetic waves which are sent to the experiment. The slave YIG oscillator drives a similar Schottky diode (the Harmonic Mixer (HM)) which mixes the harmonics of the master, coming from the experiment, with harmonics of the slave (see Figure~\ref{MVNAflow}).
 
For each harmonic there is a corresponding pair of diodes optimized for that frequency range. This optimization includes variable biasing of the diodes, the size of the diode environment and output coupling to the appropriate dimension waveguide. Coupling to mono-moded waveguide has an advantage in that it acts as a high-pass filter to help eliminate the lower, often more powerful harmonics. 

When mixed together, only harmonics of the desired rank, $N$, generate a difference frequency corresponding to that of the vector receiver, $N \times f_{\rm MHz}$. For bench-top and DC-field experiments, the vector receiver filters, amplifies and down converts the $f_{\rm MHz}$ signal to remove the other harmonics. This narrow bandwidth filtering results in a high signal to noise ratio at the expense of the time response of the of the system. To obtain the desired temporal response for use in millisecond duration pulsed fields, this latter stage must be bypassed. As a result, care must be taken to ensure that the contribution from unwanted harmonics is minimal. This is especially important when measuring photonic bandgap structures, where an adjacent harmonic (although orders of magnitude weaker) may fall outside the bandgap and thus appear with comparable intensity in the transmitted signal. 

Loading a resonator with a sample causes both a dissipative and reactive perturbation, {\it i.e.} it reduces the Q-factor and retunes the resonant frequency. In conventional (not pulsed field) experiments, field-induced changes in the sample properties can be observed in both the change in Q-factor and change in resonant frequency~\cite{MarijeJPCM,HillRSI}. This requires that the field variation is slow enough to be able to track the resonance as well as measure its amplitude. For pulsed magnetic field experiments, the frequency of the MVNA is locked to that of the resonator at zero field such that changes in amplitude of the transmitted signal contain both dissipative and reactive components. By performing magnet pulses with the measurement frequency locked slightly off resonance the relative size of the dissipative and reactive changes with field can be assessed. This painstakingly slow reconstruction of the spectra has only been performed in a limited number of cases in all of which the dissipative component has been found to dominate the measured change in cavity transmission. For example, using a moderately large sample, the EPR line from hydrated copper sulphate that causes a $-30\%$ change in cavity transmission only retunes the resonator by approximately $0.01\%$, about a $10^{{\rm th}}$ of the loaded bandwidth. 

Two methods of recording the amplitude during the pulse have been successfully employed (see the right hand side of Figure~\ref{MVNAflow}). The simplest method is to rectify the output of the vector receiver (the $N \times f_{\rm MHz}$ signal) to obtain a DC voltage that can be amplified filtered and digitized with appropriate bandwidth so as not to limit the temporal response of the measurement. This works well for resonances with a good signal to noise ratio and has the advantage of instantly yielding the transmitted amplitude. For a resonance with a poor signal to noise ratio, or in the situation where neighboring harmonics ($(N\pm1) \times f_{MHz}$) are substantially present in the vector receiver's output, the quality of the data can be improved by further mixing. By mixing the vector receiver's output with a tunable external source, a difference frequency of a few hundred kHz is obtained. This enables the AC signal to be digitized, using  a 16~bit 10~MHz GaGe card \cite{GaGe} without encountering a Nyquist sampling problem. Mixing and recording the AC signal in this manner enables further filtering both pre and post digitization to improve the signal to noise ratio. This method of data collection has the disadvantage that amplitude data are not immediately available after the magnet pulse, but require processing. In all cases, the magnetic field pulse is determined in the usual fashion~\cite{HerlachRepProgPhys} by integrating the voltage induced in a small coil of known area.

It should be noted that although in the pulsed-field experiments only the amplitude of the signal is recorded, the vector nature of the MVNA is invaluable when it is used as a diagnostic tool whilst setting up the experiment and testing new apparatus. This is especially true when measuring dielectric components where one is frequently attempting to identify a weak resonance in the presence of a large background microwave leak signal. In this case, the rapid change of transmitted phase is still present when the frequency is swept through a resonance of comparable amplitude to the background. 

\section{System performance.} \label{SSTests}
This section describes several trial examples that demonstrate the performance and versatility of the photonic bandgap resonator in pulsed magnetic fields. Resonators of both the Fabry-Perot and puck geometries have been successfully used in pulsed magnetic fields. The choice of which geometry to use is dependent upon the sample shape and dielectric properties. For example, the Electron Paramagnetic Resonance (EPR) experiments, described in Subsection~\ref{SSSEPRdata}, were performed using the puck geometry resonator, where an adequate perturbation was achieved by placing the sample radially adjacent to the puck. For the magneto-conductivity experiments, described in Subsection~\ref{SSSMagcond}, the Fabry-Perot geometry resonator was used. Although this geometry has a lower unloaded Q-factor it enables the sample to be located centrally (See Figure~\ref{DevicePic1}b allowing for a greater perturbation of the resonator resulting in an increased sensitivity. 

\subsection{Electron Paramagnetic Resonance} \label{SSSEPRdata}
EPR is observed when the microwave frequency matches the energy separation of Zeeman-split atomic energy levels whose $S_z$ or $J_z$ quantum numbers differ by $\pm1$~\cite{EPR1}.  For a spin $^1\slash_2$  system with a spectroscopic g-fator of 2, the Zeeman splitting is $\approx$~28~GHzT$^{-1}$. A measurement of a simple system of isolated spins at a frequency of 60~GHz is thus expected to reveal an EPR absorption at a magnetic field $\approx 2$~T. Comparison of EPR data from simple spin systems, for example hydrated copper sulphate, measured in conventional DC-field apparatus with data measured in pulsed fields thus provides a means of assessing the sensitivity of the pulsed-field apparatus. In such experiments, the signal to noise levels ($\approx 500$) the mass of the sample ($\approx 1$~mg) and the measured line width ($\approx 500$~G) indicate that the typical sensitivity of the pulsed-field measurement system is of the order of $2\times10^{13}$~spins~G$^{-1}$. State of the art DC-field apparatus operating in this frequency range \cite{Burghaus} achieves an EPR sensitivity of $2\times10^{7}$~spins~G$^{-1}$ with a detection time constant of 1~s.  Assuming that the signal to noise ratio in the pulsed field apparatus scales like the noise equivalent power, with the inverse square root of the detector bandwidth, the short detector time constant in pulsed fields ($<~5\times10^{-5}$~s) makes this comparison much more favorable ($2.8\times10^{10}$~spins~G$^{-1}$Hz$^{\frac{1}{2}}$ vs $2\times10^{7}$~spins~G$^{-1}$Hz$^{\frac{1}{2}}$).

In more complex systems with higher moments, exchange coupling between the spins and zero field splitting of the magnetic energy levels, even low-energy EPR lines may occur at very high magnetic fields~\cite{SrCuBOEPR}.  In systems where the crystal field anisotropy causes a significant zero-field splitting of the energy levels, a higher magnetic field is required to drive the energy levels into proximity and observe EPR. NiCl$_{2}$-thiourea is an example of a material in which the applied magnetic field drives a level crossing that gives rise to a change of ground state. This system exhibits EPR associated with the level crossing at moderately high magnetic fields \cite{RMcDNithiourea}, thus providing a more robust comparison of DC-field and pulsed field measurement schemes.

\begin{figure}
\centering
\includegraphics[height=7cm]{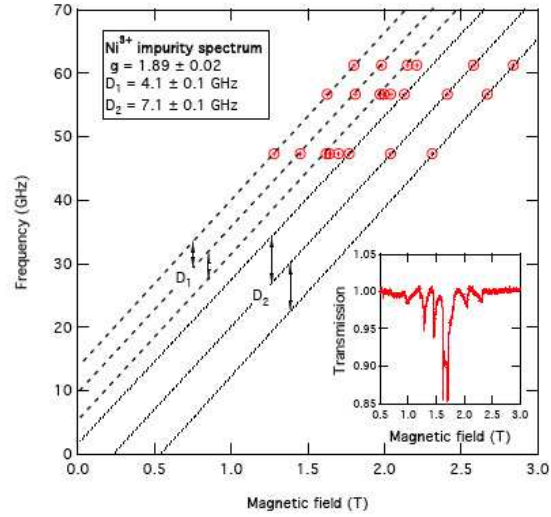}
\caption{a. The predicted (solid lines) and measured (square data points, DC field and round data points pulsed field) magnetic field/frequency dependence of the EPR lines in NiCl$_{2}$-thiourea for an applied magnetic field parallel to the {\bf c}-axis. b. The line shape of the EPR lines measured in pulsed magnetic field, at a temperature of 2.1~K. c. The line shape of the EPR lines measured in DC magnetic field, at a temperature of 1.5~K. The lines are offset vertically for clarity. Note that despite the superior Q-factor and data averaging in the DC field measurements the signal to noise ratio is similar for both data sets (color online).}
\label{EPR2}
\end{figure}

In NiCl$_{2}$-thiourea, the Nickel has a spin of one. However, the tetragonal symmetry causes a zero field splitting, $D$, of the singlet $|0>$ and triplet states $\pm|1>$ that is greater than the antiferromagnetic exchange coupling, $J$. As a result, long range magnetic order does not occur at zero field, but a magnetic field applied parallel to the c-axis causes a crossing of the spin up and singlet states, which in turn gives rise to magnetic order at finite fields. Magnetic susceptibility data~\cite{Armando} indicate an antiferromagnetic ground state below a temperature of 1~K, in the field range 2-12~T.  This suggests the following values of $g~=~2.26$, $D~=~7.6$~K and $J~=~4.5$~K. The $\Delta$S~$\pm$~1 selection rule implies that two EPR lines will be observed in the experimental frequency range: the first heading to zero frequency at 7.7~T at a rate of $-31.64$~GHzT$^{-1}$, the second emerging at 7.7~T at a rate of 31.64~GHzT$^{-1}$. This prediction is plotted as the solid lines in Figure~\ref{EPR2}a. 

DC magnetic field measurements using a conventional metallic resonator reveal the corresponding EPR lines as dips in the resonator Q-factor (see Figure~\ref{EPR2}c). Although the observed lines have significant width, at frequencies below approximately 20~GHz only one line is observed. The square data points in Figure~\ref{EPR2}a plot the frequency of the absorption dips as observed in the DC field experiments a function of field. Above 20~GHz, the gradients of the EPR lines agree with prediction. Moreover, the field at which the two lines combine is close to 7.7~T. Both of these facts suggest that the above parameters~\cite{Armando} are a not unreasonable basis for a preliminary understanding of this material's magnetic properties. The occurrence of a single EPR line below 20~GHz is not explicable by thermal broadening of isolated spin energy levels. It is hence expected to be linked to the energy scale of the spin interactions. It should also be noted that 20~GHz is equivalent to 1~K, the upper temperature of the antiferromagnetic order. Further experiments are underway to investigate the effect of the magnetic order upon the EPR spectra at low temperature. Figure~\ref{EPR2}b shows the EPR spectrum measured using the photonic bandgap resonator in pulsed magnetic fields at a temperature of 2.1~K. The circular data points that extend Figure~\ref{EPR2}a to higher frequency and field are extracted from the pulsed field data. 

\begin{figure}
\centering
\includegraphics[height=7cm]{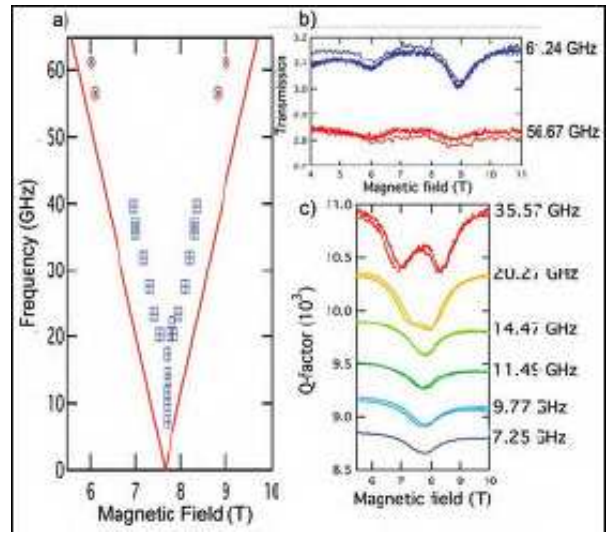}
\caption{The frequency magnetic field dependence of the EPR features observed at low fields in NiCl$_{2}$-thiourea. The inset shows the corresponding spectrum measured at a frequency of 47.1~GHz and a temperature of 2.1~K (color online). }
\label{EPR3}
\end{figure}

The low magnetic field part of the NiCl$_{2}$-thiourea spectrum contains multiple sharp features attributed to Ni$^{3+}$ impurities (see the inset to Figure~\ref{EPR3}) and thus can be used as a measure of sample quality. The assignment as Ni$^{3+}$ EPR is based on the fact that the lines appear in groups of three corresponding to the allowed transitions ($\Delta$S = $\pm$1) for a spin $^3/_2$ system split by the crystal field. At least two impurity sites can be distinguished owing to differing zero field splittings arising from different ligand fields (see the dashed and dotted lines in Figure~\ref{EPR3}).  These EPR lines all have the same g-factor of 1.89 $\pm$ 0.02, larger than the unquenched Ni$^{3+}$ spectroscopic g-factor of 1$^1$/$_3$ and smaller that the fully quenched g-factor of 2, again consistent with the impurity interpretation. Comparing the spectral weight of the impurity features with that obtained from a sample with a known number of spins it is possible to estimate the concentration of impurities in this sample, which is of the order of one in $10^3 - 10^4$ unit cells. 

\subsection{High-frequency magneto-conductivity measurements} \label{SSSMagcond}

The other class of experiment performed with the photonic bandgap resonator is the measurement of metallic systems in which the GHz-frequency conductivity is dominated by the behavior of electronic states at or close to the Fermi energy.

\begin{figure}
\centering
\includegraphics[height=7.8cm]{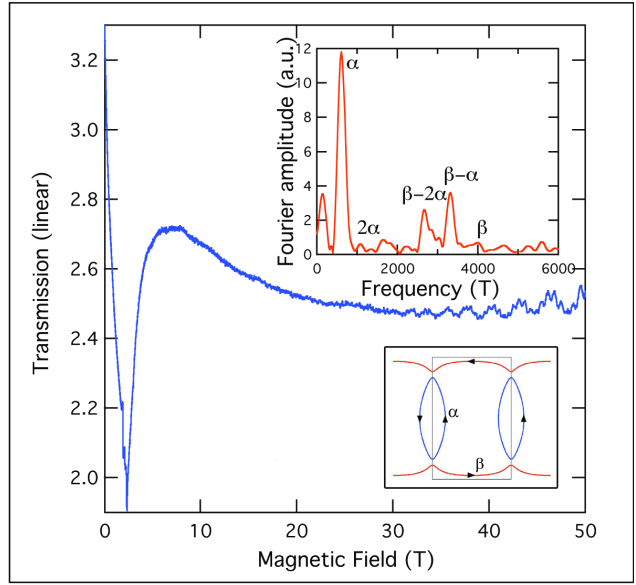}
\caption{The GHz-frequency magneto-conductivity of $\kappa$-(BEDT-TTF)$_{2}$Cu(SCN)$_{2}$ measured at a temperature of 2.1~K and a frequency of 64~GHz. Note the Josephson plasma resonance at low field and the quantum oscillations at higher fields. The data in this figure was acquired during a single magnet pulse. The insets show the Fourier spectrum of the quantum oscillations and the associated Fermi-surface orbits (color online). } \label{QoscFig}
\end{figure}

In metals the magnetic field quantizes the orbital angular momentum, splitting the continuous electron dispersion into discrete Landau levels. With increasing magnetic field, both the energy separation and degeneracy of the Landau levels increases, which causes an oscillation in the density of states at the Fermi energy, periodic in inverse magnetic field. This effect is observed as magnetic quantum oscillations in the sample conductivity, magnetization and other properties~\cite{Shoenburg}. An example of the this kind measurement performed using the Photonic bandgap resonator is the magneto-conductivity of  the quasi-two-dimensional crystalline organic superconductor $\kappa$-(BEDT-TTF)$_{2}$Cu(SCN)$_{2}$~\cite{SingletonRepProgPhys}. The Fermi surface cross-section of this material is shown in the lower inset of Figure~\ref{QoscFig}.

The data in Figure~\ref{QoscFig} were measured during a {\it single} magnetic-field pulse with a rise time of 8~ms using a frequency of 64~GHz and a sample temperature of 2.1~K. Under these conditions, quantum oscillations corresponding to the $\alpha$ and $\beta$ orbits about the(plus Stark quantum interference between the two) are clearly visible (see the upper inset of Figure~\ref{QoscFig}).  The dip in transmission at low field is a Josephson plasma resonance, which occurs close to the material's superconducting upper critical field~\cite{JPL}. 

At this point it is useful to consider the factors that limit the observation of magnetic quantum oscillations. First, the width of the thermally-broadened tail of the Fermi-Dirac distribution function should usually be less than the energy separation of the Landau levels~\cite{Shoenburg}. This corresponds roughly to the condition $\hbar \omega_{\rm c}  \geq k_{\rm B}T$, where $\omega_{\rm c}$ is the cyclotron frequency. In the case of $\kappa$-(BEDT-TTF)$_2$Cu(NCS)$_2$, the effective mass of the alpha orbit is $3.5 \pm 0.1m_{\rm e}$ and that of the beta orbit $6.5 \pm 0.5m_{\rm e}$, so that the observation of quantum oscillations at around 2 K is only possible at the high fields shown in Figure~\ref{QoscFig}~\cite{MarijeJPCM,SingletonRepProgPhys}.
In a dirty or disordered material, the quasiparticle scattering rate $\tau^{-1}$ produces a further restriction: for oscillations to be observable, the separation of the Landau levels,  $\hbar \omega_{\rm c}$ must be greater than or of the order of their broadening $\sim \tau^{-1}$. For a given effective mass, temperature and scattering rate, this translates to a quantum-oscillation amplitude that grows exponentially with magnetic field via the so called Dingle factor~\cite{Shoenburg}. One can therefore use extremely high magnetic fields to probe the band structure of dirty or disordered materials. However, this requires instrumentation that can cope with the associated extreme rates of change of magnetic field found in pulsed magnets. Here we have demonstrated a technique that unlike conventional conductivity and magnetization measurements~\cite{PaulAuZn} is not limited by the rate of change of magnetic flux. Hence, it could find application in extremely high pulsed magnetic fields such as those provided by single-turn magnets in which $\frac{dB}{dt} \sim 10^{8}$~Ts$^{-1}$ \cite{JSstp}.

\section{Conclusion.}
\label{summary}
We have discussed  instrumentation for making microwave-frequency measurements in pulsed magnetic fields and illustrated the performance of the apparatus with several examples. 
The major component of this system is the non-metallic microwave resonator, in which the radiation is confined by the use of periodic dielectric arrays. The dimension of the resonator and period of array is chosen such that the resonant frequency coincides with the photonic bandgap of the structure so as to form an efficient, high Q-factor resonator. 

In addition to operation in environments with extreme rates of change of magnetic field the non-metallic nature of this resonator lends itself to other applications, for example, operation in remote or chemically reactive environments where degradation of the surface of a metallic resonator would compromise performance. The performance of metallic resonators is very sensitive to surface cleanliness and oxidation; by contrast the photonic bandgap structures are chemically inert and distributed reflectors, potentially reducing the necessity of maintenance such as surface cleaning.  Another potential advantage over a metallic system is the degree of temperature stability. Unlike metallic resonators, where the conductivity of the metal and hence the Q-factor of the resonator is a strong function of temperature \cite{Pozar}, the dielectric properties of the photonic bandgap structures can be chosen to have a relatively small temperature dependence. Along similar lines this design of resonator may be suitable for high power microwave applications. At high microwave powers the associated ohmic heating due to the induced screening currents limits the Q-factor of metallic cavities \cite{Pozar}. The distributed nature of a photonic bandgap resonator ensures that the power dissipated within is done so over its entire volume, not just at the surface. As a result the heat associated with a pulse of microwave power will lead to a smaller change in temperature. A photonic bandgap resonator could thus potentially be designed to operate at greater peak powers than a conventional metallic system. This may prove advantageous for fast pulsed systems where the cooling of a metallic resonator is limited by thermal time constants.

\section{Acknowledgments.}
This work is supported by US Department of Energy (DOE) grants LDRD20030084DR 
and LDRD2004009ER and was performed under the auspices 
of the National Science Foundation, the DOE and the State of Florida.
We thank Arzhang Ardavan, Phillipe Goy, Steve Hill, Albert Migliori, Alessandro Narduzzo and Sidney Self for stimulating discussions.

\newpage

\begin{table}[	ht]
\begin{center}
\begin{tabular}{cc} \hline \hline
Mode & Frequency (GHz) \\
  \hline\hline
  HEM$_{111}$ & 51.3\\
   TE$_{011}$ & 56.1\\
    HEM$_{211}$ & 57.4\\
     TM$_{011}$ & 57.9\\
      HEM$_{121}$ & 63.1\\
      HEM$_{311}$ & 65.3\\
      HEM$_{131}$ & 66.5\\
      HEM$_{221}$ & 70.4\\
      \hline
\end{tabular}
\caption{The calculated mode frequencies for the puck geometry resonator with a 3~mm raduis, 2~mm long cylindrical Rexolite puck.}
\end{center}
\label{modetable}
\end{table} 

\newpage

\end{document}